\documentclass[letter]{aa}

\usepackage{graphicx}
\usepackage{txfonts}
                                
\usepackage[colorlinks=true, citecolor=blue, linkcolor=blue, urlcolor=blue]{hyperref}

\begin{document}

   \title{Coexistence of Chromatic Flares and an Achromatic QPO in the Gamma-ray Blazar PG 1553+113}


    \titlerunning{Chromatic Flares and an Achromatic QPO in PG 1553+113}

%

\author{Elena Madero\inst{1}\fnmsep\thanks{emadero@ucm.es}
        \and Alberto Dom\'inguez\inst{1,2}\fnmsep\thanks{alberto.d@ucm.es}
        }

\institute{$^1$Department of EMFTEL, Universidad Complutense de Madrid, E-28040 Madrid, Spain \\
            $^2$Instituto de Física de Partículas y del Cosmos (IPARCOS), Universidad Complutense de Madrid, E-28040 Madrid, Spain}

   \date{Received January 26, 2026}

 
\abstract
{The physical origin of quasi-periodic oscillations (QPOs) in blazars remains debated, with geometric and plasma-driven scenarios as the main competing interpretations. Discriminating between them requires probing variability beyond flux periodicity.}
{We study the spectral evolution of the BL Lac object PG 1553+113 over its $\sim$2.2-year gamma-ray QPO cycle to constrain the mechanism driving the oscillation. In particular, we test whether the variability is chromatic (coupled to spectral changes) or achromatic (independent of spectral shape), allowing us to distinguish between plasma-driven and geometric scenarios.}
{We analyze 17 years of \textit{Fermi}-LAT data (2008--2025) with 30-day binning. To mitigate red-noise effects, we apply Singular Spectrum Analysis (SSA) to remove slow baseline trends and use a Block Bootstrap approach to quantify correlations between photon flux and photon index while preserving temporal dependence.}
{We find a robust softer-when-brighter chromatic trend, atypical for high-synchrotron-peaked blazars such as PG 1553+113 and which, based on our analysis, physically corresponds to softer-when-flaring episodes, that persists after detrending and accounting for temporal autocorrelation. In contrast, no significant correlation is detected between the photon index and the QPO phase, indicating that the periodic modulation is effectively achromatic.}
{The coexistence of plasma-driven chromatic flares and an achromatic QPO disfavors scenarios in which the periodicity arises from intrinsic jet processes. Instead, the results support a geometric origin for the QPO modulation, such as jet precession, where Doppler-factor variations modulate the flux without altering the intrinsic particle energy distribution.}

\keywords{galaxies: active --
                galaxies: jets --
                gamma rays: galaxies --
                radiation mechanisms: non-thermal --
                methods: statistical --
                BL Lacertae objects: individual: PG 1553+113
               }

   \maketitle
\nolinenumbers

\section{Introduction}

Blazars are radio-loud active galactic nuclei (AGNs) with relativistic jets aligned with the line of sight and show extreme variability across the electromagnetic spectrum \citep[e.g.,][]{Urry1995, Abdo2010}. Although this variability is mainly stochastic, some sources display quasi-periodic oscillations (QPOs) in their light curves, whose physical origin remains widely debated in high-energy astrophysics \citep[e.g.,][]{Sandrinelli2016, Jorstad2022}.

The BL Lacertae object PG 1553+113, classified as a High-Synchrotron Peaked (HSP) blazar \citep[e.g.,][]{NievasRosillo2022, Lainez2025, Dinesh2025}, represents a benchmark case for such studies. \citet{Ackermann2015} reported a compelling QPO in its gamma-ray flux with a period of $\approx 2.2$ years, a result that has been subsequently confirmed and extended by later analyses \citep{Tavani2018, Penil2020, Penil2024, Abdollahi2024, Rico2025}. Note that continuous wavelet transforms may yield lower QPO significance due to high trial factors \citep{Ren2023}. The persistence of this oscillation has fueled the hypothesis that PG 1553+113 hosts a supermassive binary black hole (SMBBH) system \citep[e.g.,][]{Sobacchi2017, Agazie2024}, where the periodicity arises from orbital interaction or hydromagnetic instabilities. Alternative scenarios propose geometric origins, such as jet precession \citep[e.g.,][]{Rieger2004, Liska2018}, which could itself be dynamically driven by a binary companion \citep{Begelman1980}.

Distinguishing whether the QPO is driven by intrinsic plasma mechanisms (e.g., periodic accretion changes or shocks) or geometric effects (e.g., a binary-driven precessing jet) requires a robust investigation of the spectral variability. Standard correlations between flux and photon index, such as the softer-when-brighter or harder-when-brighter trends, can trace the underlying particle acceleration processes \citep[e.g.,][]{Fossati1998, Bhatta2016}. However, blazar light curves are dominated by red noise, which can introduce spurious correlations if not properly treated \citep[e.g.,][]{Vaughan2003}. Recent works using SSA have demonstrated the importance of decomposing blazar light curves to separate intrinsic variability and noise from baseline trends \citep[e.g.,][]{Penil2025,Rico2025}.

In this Letter, we constrain the physical driver of the QPO in PG 1553+113 by analyzing 17 years of \textit{Fermi}-LAT data. We employ a robust statistical framework, combining SSA detrending with Block Bootstrap resampling \citep[e.g.,][]{Kunsch1989}, to test for coupling between the photon index and the flux, as well as the QPO phase.

\section{Theoretical Background}

\subsection{Emission Models, photon index, and Variability}

The broadband spectral energy distribution (SED) of HSP blazars such as PG 1553+113 is typically described by the one-zone Synchrotron Self-Compton (SSC) model \citep[e.g.,][]{Katarzynski2001, Aleksic2015}. In this framework, the emission originates from a spherical region of plasma within the relativistic jet, characterized by a magnetic field $B$ and moving with a Doppler factor $\delta$. The low-energy peak is produced by synchrotron radiation from ultra-relativistic electrons, while the high-energy gamma-ray peak arises from inverse Compton scattering of these same synchrotron photons by the parent electron population \citep[e.g.,][]{RybickiLightman1979}.

The observed gamma-ray photon index, $\Gamma$, derived from the power-law approximation $dN/dE \propto E^{-\Gamma}$, serves as a proxy for the underlying electron energy distribution index, $p$. Under standard radiative cooling conditions, these quantities are related by $\Gamma = (p+1)/2$ \citep[e.g.,][]{Abdollahi2020}. Variations in $\Gamma$ therefore probe the microphysics of the emitting region, reflecting changes in the balance between particle acceleration and radiative cooling rates.

Correlations between the gamma-ray flux ($F_{\gamma}$) and photon index ($\Gamma$) provide additional information on the physical drivers of variability \citep[e.g.,][]{Abdo2010}. A harder-when-brighter behavior, characterized by a decrease in $\Gamma$ as the flux increases, is commonly interpreted as evidence for the injection of fresh high-energy electrons through efficient acceleration processes, such as shocks, and is often associated with flaring states \citep[e.g.,][]{Sambruna2000}. In contrast, a softer-when-brighter trend, in which $\Gamma$ increases with flux, may indicate variability driven primarily by changes in particle density or enhanced radiative cooling rather than by spectral hardening \citep[e.g.,][]{Abdo2010}.

\subsection{Geometric vs. Plasma-driven Modulation}
Physical models for QPOs can be broadly categorized by their prediction of chromatic versus achromatic variability. Intrinsic plasma-driven models, such as regular accretion changes induced by a binary companion \citep{Valtonen2009} or gravitational 
stresses \citep{Tavani2018}, predict phase-dependent spectral changes 
in the electron distribution, leading to chromatic signatures like 
periodic spectral hardening \citep[e.g.,][]{Sobacchi2017}. Conversely, geometric models, such as jet precession, attribute the variability to periodic changes in the Doppler factor $\delta$. Since the bolometric flux scales as $F_{\gamma} \propto \delta^4$ while the spectral shape remains largely achromatic in the comoving frame \citep[e.g.,][]{Rieger2004, Malik2025}, geometric origins are expected to produce flux modulation with little to no dependence of the photon index on the QPO phase. Finding an achromatic QPO would thus disfavor accretion- or shock-driven periodicities, while remaining fully compatible with a binary-driven precessing jet.

\section{Methodology}
\label{sec:methods}

\subsection{Data and Source Selection}

The data for this analysis were obtained from the \textit{Fermi} Large Area Telescope (LAT), the primary instrument on board the \textit{Fermi Gamma-ray Space Telescope} \citep{Atwood2009}.

We accessed the data through the \textit{Fermi}-LAT Light Curve Repository\footnote{\url{https://fermi.gsfc.nasa.gov/ssc/data/access/lat/LightCurveRepository/}} \citep[LCR,][]{Abdollahi2023}, a public database providing flux and spectral information for sources in the 4FGL catalog \citep{Abdollahi2020}, whose automated products for PG 1553+113 have been validated against targeted manual reductions \citep[e.g.,][]{Penil2024, Abdollahi2024, Rico2025}. The dataset for PG 1553+113 spans a 17-year interval, covering the period from August 2008 to September 2025 (approximately MJD 54,600 to MJD 61,000).

To characterize this multi-year variability and ensure sufficient statistics for spectral fitting, we selected a time binning of 30 days. This integration time is significantly shorter than the QPO period, allowing for a detailed phase-resolved analysis, while effectively smoothing out short-term stochastic fluctuations that could mask the periodic modulation. The analysis covers the standard \textit{Fermi}-LAT energy range from 100 MeV to 100 GeV, following the standard LCR methodology.

We utilized the ``Free Index'' fitting option provided by the LCR, which allows the photon index $\Gamma$ to vary in each time bin. To ensure data quality, we applied a Test Statistic (TS) threshold of $TS \geq 4$ (approximately equivalent to a $2\sigma$ detection) for each bin. Time intervals failing this criterion, which often correspond to high background noise or low source flux states, were excluded to avoid introducing spurious outliers in the correlation analysis.

The final dataset consists of simultaneous measurements of $F_{\gamma}$ and $\Gamma$, along with their respective uncertainties, serving as the input for the SSA and correlation routines described below.

\subsection{Correlation Analysis and Treatment of Red Noise}

We quantified the relationship between flux and photon index using two independent metrics: the Pearson correlation coefficient ($r$), which evaluates linear dependence, and the Spearman rank correlation coefficient ($\rho$), which assesses monotonic relationships and is less sensitive to outliers \citep{Walpole2012}.

To account for experimental uncertainties in the \textit{Fermi}-LAT data, we employed a Monte Carlo bootstrap approach. We generated $N=2000$ synthetic datasets by perturbing both flux and photon index values within their Gaussian observational uncertainties. Accordingly, the analysis is anchored to the results obtained from the unperturbed data, which are adopted as the nominal values, while the $1\sigma$ and $2\sigma$ uncertainties are inferred from the spread of the corresponding bootstrap distributions in order to mitigate attenuation bias associated with the randomization \citep[e.g.,][]{Peterson1998}.

Standard bootstrap methods assume data independence, an assumption violated by the temporal autocorrelation intrinsic to blazar variability. To preserve the time-dependent structure of the light curves and mitigate red-noise effects, we adopt a block bootstrap approach \citep[e.g.,][]{Kunsch1989}. The 17-year light curve is divided into consecutive blocks of size $k=4$ bins (corresponding to 120 days), a scale chosen to retain local temporal correlations while still allowing adequate randomization \citep[e.g.,][]{Politis1994}. Synthetic light curves are then constructed by resampling these blocks with replacement, ensuring that the resulting confidence intervals for $r$ and $\rho$ remain robust against red-noise artifacts. To verify that our results are not sensitive to this specific choice, we perform a consistency check by varying the block size over the range $k \in [2, 6]$.

\subsection{Detrending via Singular Spectrum Analysis}
Baseline trends in blazar emission can introduce artificial correlations that mask shorter-term variability. To isolate the QPO and flaring components from this slow evolution, we applied SSA, a non-parametric spectral estimation technique well suited for non-stationary astrophysical time series; we refer the reader to \citet{Rico2025} for complete SSA configuration details.

The SSA decomposition separates the light curve into a slowly varying trend, oscillatory components (including the QPO), and noise. We subtracted the reconstructed trend component from the original light curve to obtain “detrended” flux values following the method described by \citet{Rico2025}. This trend may be part of a longer, $\sim$22 yr QPO \citep{Adhikari2024}. We then repeated the correlation analysis on these detrended data to test whether the spectral trends persist in the absence of secular baseline variations.

\subsection{Phase-Dependent Spectral Analysis}
Finally, to test the hypothesis of geometric precession versus binary interaction, we folded the photon index data over the reported QPO period of PG 1553+113 \citep[$P_{\rm QPO} \approx 2.2$ years, e.g.,][]{Ackermann2015, Abdollahi2024, Penil2025}. We assigned a phase $\phi \in [0,1]$ to each 30-day bin and analyzed the dependence of $\Gamma$ on $\phi$. A significant correlation here would imply that the spectral hardness is modulated by the mechanism driving the periodicity, as predicted by binary shock models.

\begin{figure}
\centering
\includegraphics[width=\columnwidth]{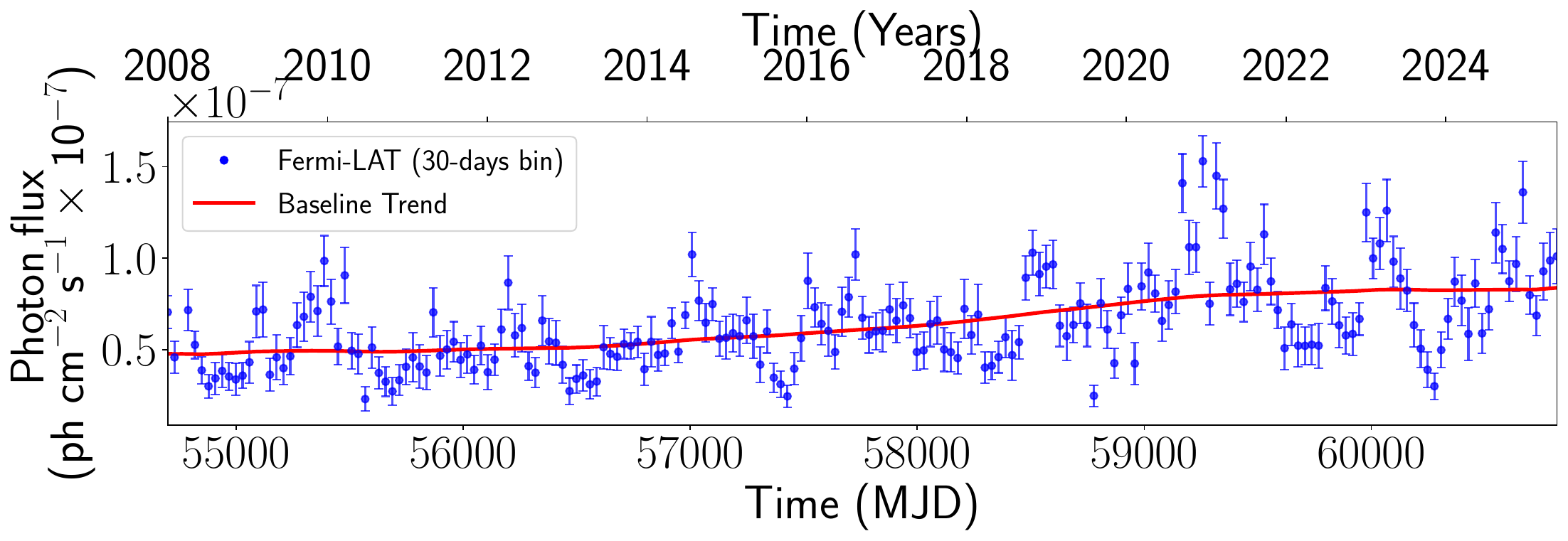}
\caption{Gamma-ray light curve of PG 1553+113 (30-day bins, 2008--2025): Fermi-LAT flux (blue circles, $1\sigma$ errors) and the SSA baseline (red line) used for detrending.}
\label{Fig:LightCurve}
\end{figure}

\begin{figure*}
\centering
\includegraphics[width=7cm]{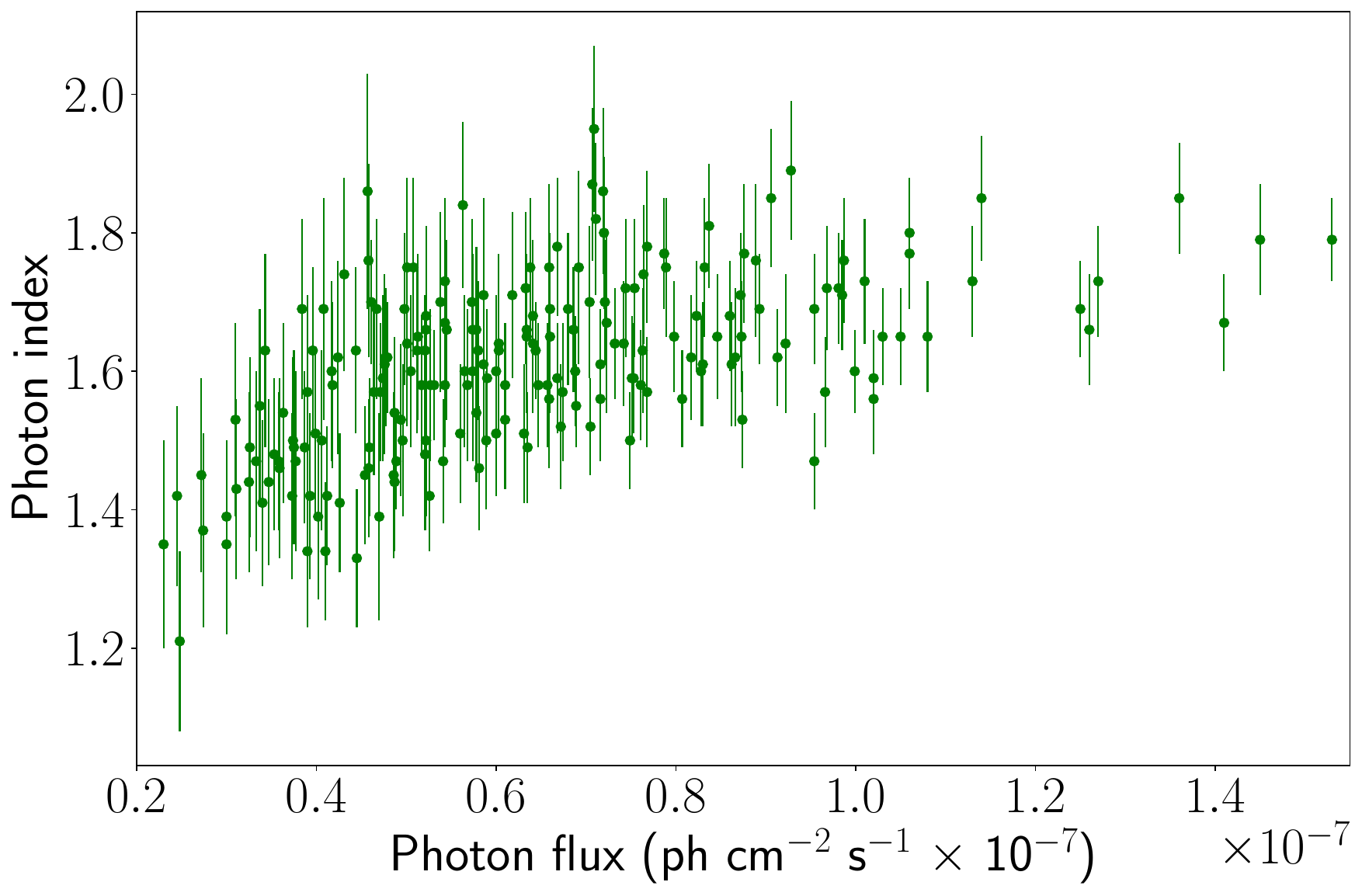}
\includegraphics[width=7cm]{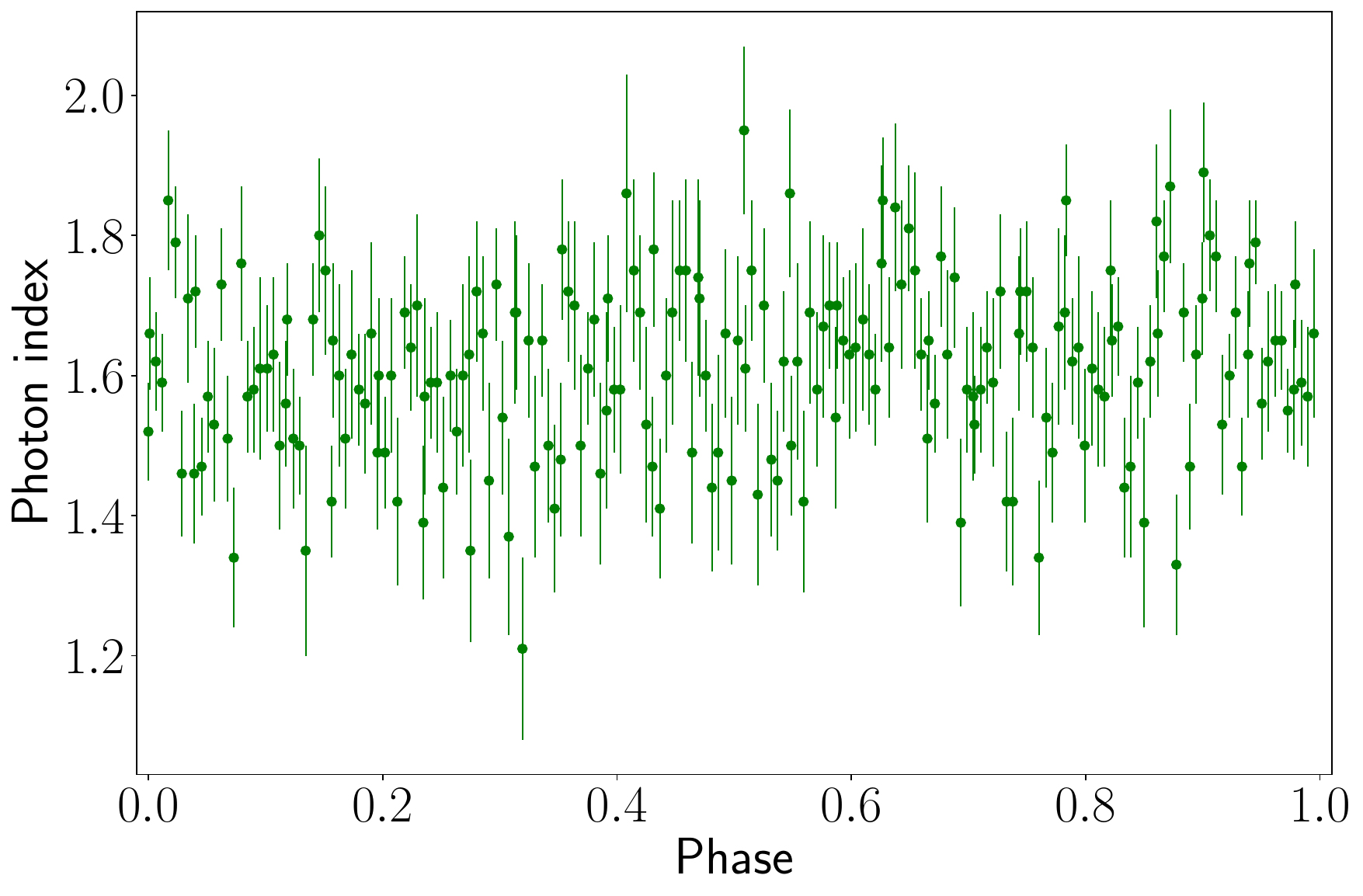}
\caption{(Left) Photon index vs. \textit{Fermi}-LAT flux, showing the correlation reported in Table~\ref{table:1} and the weighted linear fit $\Gamma = (0.28 \pm 0.03)(F_{\gamma} \times 10^7) + (1.43 \pm 0.02)$. (Right) Photon index vs. QPO phase, showing no correlation ($\rho \approx 0.04$, $p = 0.57$).}
\label{Fig:Correlation}
\end{figure*}

\section{Results and Discussion}
\subsection{Flux and Spectral Variability}
\label{sec:flux_spectral}

The 17-year light curve of PG 1553+113 (MJD 54,600--61,000) is shown in Fig. \ref{Fig:LightCurve}. The source shows significant variability on multiple timescales, including short-term flares and the well-reported multi-year modulation \citep[e.g.,][]{Ackermann2015}. SSA successfully isolates the baseline trend (Fig. \ref{Fig:LightCurve}), capturing the slow evolution of the baseline flux without distorting the short-term flaring activity \citep{Rico2025}.

We investigated the relationship between $F_{\gamma}$ and $\Gamma$. As shown in Fig. \ref{Fig:Correlation}, we observe a clear positive correlation, indicating a chromatic softer-when-brighter behavior, contrasting with the harder-when-brighter trend typical of HSPs \citep{Abdo2010}. The Pearson ($r$) and Spearman ($\rho$) correlation coefficients for the original dataset are $r = 0.57 \pm 0.06$ and $\rho = 0.59 \pm 0.06$.

A critical concern in blazar studies is that red noise can mimic apparent correlations. To assess the robustness of our results, we compared the correlation coefficients derived from the original data with those obtained after detrending the light curve through removal of the SSA trend, using both standard bootstrap and Block Bootstrap methods. The corresponding results are summarized in Table~\ref{table:1}. The correlation persists after detrending, with coefficients of $r \approx 0.51$ and $\rho \approx 0.50$. In all cases, the $2\sigma$ confidence intervals overlap and exclude zero, providing a statistically robust confirmation that the observed relationship is not a product of chance or red-noise fluctuations. Even when varying the block size over the range $k \in [2, 6]$, the resulting 95\% confidence interval remains $[0.42, 0.73]$, strictly excluding zero and confirming the robustness of the correlation. Importantly, the softer-when-brighter spectral trend is already present in the original light curve and remains unchanged after removal of the secular baseline component via SSA, indicating that the correlation is not driven by slow baseline evolution but reflects an intrinsic property of the stochastic variability.

\begin{table}[ht!]
\caption{Correlation coefficients and 95\% confidence intervals.}
\label{table:1}
\centering
\setlength{\tabcolsep}{3pt} 
\begin{tabular}{l c c c c}
\hline\hline
Method & $r \pm 1\sigma$ & 95\% CI ($r$) & $\rho \pm 1\sigma$ & 95\% CI ($\rho$) \\
\hline
Obs. (Standard)  & $0.56 \pm 0.06$ & $[0.44, 0.67]$ & $0.59 \pm 0.06$ & $[0.46, 0.70]$ \\
Detr. (Standard) & $0.51 \pm 0.06$ & $[0.38, 0.62]$ & $0.50 \pm 0.07$ & $[0.36, 0.62]$ \\
Obs. (Block)     & $0.57 \pm 0.06$ & $[0.45, 0.68]$ & $0.59 \pm 0.06$ & $[0.46, 0.71]$ \\
Detr. (Block)    & $0.51 \pm 0.06$ & $[0.39, 0.63]$ & $0.50 \pm 0.07$ & $[0.36, 0.64]$ \\
\hline
\end{tabular}
\end{table}



\subsection{Phase-Resolved Spectral Behavior}
Fig. \ref{Fig:Correlation} shows the photon index folded over the $\sim$2.2-year QPO period. No phase-dependent spectral modulation is detected. The Spearman rank coefficient between phase and index is negligible ($\rho = 0.04$) with a p-value of $0.57$, indicating statistical independence.

This result clarifies the flux–spectral behavior discussed in Section \ref{sec:flux_spectral}. Although a clear chromatic trend is present in the flux variations, it disappears when the data are folded on the QPO phase. The softer-when-brighter behavior therefore arises from fast variability processes, on timescales shorter than the 30-day binning, rather than from the periodic modulation. Rapid spectral evolution is governed by plasma-related mechanisms, while the QPO modulates the flux without introducing additional spectral changes, implying an effectively achromatic origin.

In binary scenarios involving disk interactions or periodic shocks, spectral hardening is expected to coincide with the flux maximum \citep[e.g.,][]{Sobacchi2017}. The absence of phase-locked spectral evolution disfavors plasma-driven binary shock models and instead supports a geometric origin for the QPO variability. In a jet precession scenario, the modulation is produced by periodic changes in the Doppler factor $\delta$. Since the observed flux scales as $F \propto \delta^p$ (with $p > 3$) while the photon index $\Gamma$ remains invariant in the comoving frame, precession naturally produces strong flux variations with minimal spectral impact \citep[e.g.,][]{Rieger2004}. The data are therefore consistent with plasma-driven spectral variability superimposed on a geometrically modulated baseline.

For HSP blazars, the measured correlation coefficients ($r \approx 0.57$, $\rho \approx 0.59$) indicate a genuine physical coupling given the limited dynamical range of the photon index in these sources. This correlation persists after removing secular trends via SSA detrending ($\rho \approx 0.50 \pm 0.07$) and remains significant at the $2\sigma$ level when temporal autocorrelation is accounted for using Block Bootstrap resampling. This behavior contrasts with the near-zero correlation between the photon index and the QPO phase, reinforcing the conclusion that fast variability is driven by intrinsic plasma processes, while the QPO is effectively achromatic and consistent with a geometric origin such as jet precession. This interpretation does not exclude the presence of a SMBBH, which could naturally induce precession without producing phase-locked spectral signatures.

The coincidence of high-flux, soft-index states reveals a robust softer-when-flaring behavior, indicating that individual flares are intrinsically softer rather than simply reflecting a global flux–index correlation. These episodes likely arise from stochastic plasma processes within the precessing jet, such as plasmoid motion \citep[e.g.,][]{Sobacchi2017} or magnetic reconnection, superimposed on geometric precession. Doppler boosting at QPO maxima amplifies concurrent flares, naturally producing the observed softer-when-brighter trend as the macroscopic imprint of these intrinsically softer flaring events over an otherwise achromatic QPO modulation.

\section{Summary and Conclusions}

In this Letter, we presented a rigorous spectral analysis of the blazar PG 1553+113 using 17 years of \textit{Fermi}-LAT data (2008--2025). Our primary goal was to discriminate between plasma-driven and geometric physical scenarios for the reported $\sim$2.2-year QPO by investigating the coupling between the photon index and the flux modulation. To ensure the validity of our results against the characteristic red noise of blazar light curves, we employed SSA for detrending and a Block Bootstrap method for uncertainty estimation.

Our main findings are summarized as follows: (1) We confirmed a statistically significant positive correlation between the photon flux and the photon index ($\rho = 0.59 \pm 0.06$). This softer-when-brighter behavior, often observed in flat-spectrum radio quasars but rare in HSP blazars, suggests that the short-term variability is driven by mechanisms where radiative cooling or particle density variations dominate over particle acceleration during high states. Crucially, this correlation remains stable ($\rho \approx 0.5$) even after removing baseline trends and accounting for temporal autocorrelation via block bootstrapping. (2) We found no significant correlation between the photon index and the phase of the QPO (p-value = 0.57, $\rho = 0.04$). While the total flux is strongly modulated over the $\sim$2.2-year cycle, the spectral shape of the gamma-ray emission does not vary systematically with the oscillation phase. (3) The absence of phase-dependent spectral modulation provides a strong constraint on the physical origin of the QPO. Models involving intrinsic periodic instabilities or binary black hole disk interactions are expected to induce spectral changes (e.g., hardening due to shocks) coupled to the flux periodicity. Conversely, our results are fully consistent with a geometric origin, such as jet precession. In this scenario, the periodic variation in the Doppler factor modulates the observed gamma-ray flux achromatically without altering the intrinsic electron distribution of the emitting region.

We conclude that PG 1553+113 shows a dual variability nature. Fast, stochastic softer-when-brighter chromatic variability is driven by internal plasma processes, while an achromatic QPO arises from a separate, large-scale geometric precession that modulates the observed flux on multi-year timescales. Future work should extend this methodology to the broader population of QPO blazar candidates, and to analogous multiwavelength studies, to assess whether this geometric signature is a common feature of quasi-periodic gamma-ray sources.

\section*{Data availability}
Data are available in the \textit{Fermi}-LAT Light Curve Repository.


%

\end{document}